\documentclass[useAMS,usenatbib]{mn2e}
\usepackage{psfig,rotating}
\usepackage{graphicx,color}
\usepackage{dcolumn}
\usepackage{natbib}
\DeclareMathAlphabet{\mathsc}{OT1}{cmr}{m}{sc}
\def\testbx{bx}%
\DeclareRobustCommand{\ion}[2]{%
\relax\ifmmode
\ifx\testbx\f@series
{\mathbf{#1\,\mathsc{#2}}}\else
{\mathrm{#1\,\mathsc{#2}}}\fi
\else\textup{#1\,{\mdseries\textsc{#2}}}%
\fi}
\def\h2{\ensuremath{\rm H_2}}

\def\kms{km\,s$^{-1}$}

\newcommand{\be}{\begin{equation}}
\newcommand{\en}{\end{equation}}

\def\zabs{$z_{\rm abs}$}
\def\zem{$z_{\rm em}$~}

\def\hi{H~{\sc i}}

\def\kms{km~s$^{-1}$}
\def\zabs{$z_{\rm abs}$}
\def\zem{$z_{\rm em}$~}

\def\hi{H~{\sc i}~}

\def\si2{Si~{\sc ii}~}

\def\halpha{H~{\sc $\alpha$}~}
\def\hbeta{H~{\sc $\beta$}~}

\def\kms{km~s$^{-1}$}

\title[Low-z merging galaxies]
{Circumnuclear and infalling H~{\sc i} gas in a merging galaxy pair at $z~=~0.123$
}
\author[Srianand et al.] {R. Srianand$^{1}$\thanks{E-mail:anand@iucaa.ernet.in}, 
N. Gupta$^{1}$, E.  Momjian$^2$, and M. Vivek$^3$  
\\
$^{1}$ IUCAA, Postbag 4, Ganeshkhind, Pune 411007, India \\
$^{2}$ National Radio Astronomy Observatory, 1003 Lopezville Road, Socorro, NM 87801, USA
 \\
$^3$ Department of Physics and Astronomy, University of Utah, Salt Lake City, UT 84112, USA\\
}
\begin{document}

\date{Accepted. Received; in original form }

\pagerange{\pageref{firstpage}--\pageref{lastpage}} \pubyear{2014}

\maketitle

\label{firstpage}

%%%%%%%%%%%%%%%%%%%%%%%%%%%%
\begin{abstract}
Using long-slit optical spectra obtained with the 2-m telescope at IUCAA 
Girawali Observatory, we show that the radio source J094221.98+062335.2 
($z=0.123$) is associated with a galaxy pair undergoing a major merger. 
Its companion is a normal star-forming galaxy infalling with a velocity of 185 \kms
at a projected separation of 4.8\,kpc. Using the Westerbork Synthesis Radio Telescope (WSRT) 
and Giant Metrewave Radio Telescope (GMRT) we detect a strong H~{\sc i} 21-cm absorption 
at the systemic redshift of the radio galaxy with $N$(H~{\sc i})$\sim 9\times10^{21}$ cm$^{-2}$ 
for an assumed spin-temperature of 100 K.
Such a strong H~{\sc i} 21-cm absorption is rare and has been seen only in
a few compact radio sources associated with similar merging galaxy pairs. 
Milliarcsecond resolution Very Long Baseline Array (VLBA) observations resolve 
the radio source into 
a compact symmetric object with the hotspot separation of 89\,pc.  
The 21-cm absorption is detected in the VLBA spectra towards both the 
radio lobes albeit with a strong 
optical depth gradient.  We show that the strong 21-cm absorption is 
consistent with it being arising from  a clumpy circumnuclear disk/torus.
We also detect two weaker absorption lines redshifted with respect 
to the radio source in the WSRT/GMRT spectrum. They probably
represent cold (i.e. $T\le 10^4$K) H~{\sc i} gas falling into the radio source.
The presence of high concentration of
H~{\sc i} gas in the circumnuclear regions and signature of infalling cold gas allows 
us to conjecture that the young radio source may have been triggered by the gas 
infall due to the ongoing merger. 
\end{abstract} 
%
%%%%%%%%%%%%%%%%%%%%%%

%%%%%%%%%%%%%%%%%%%%%%%%%%%
\section{Introduction}
%%%%%%%%%%%%%%%%%%%%%%%%%%%%

It is now well recognised that active galactic nuclei (AGNs) play a vital role 
in the formation and evolution of their host galaxies 
\citep[e.g.][]{Silk98, Ferrarese00, Fabian12}.
The so-called AGN feedback through winds from the accretion disk or by the relativistic 
radio jets can deposit large amounts of energy and momentum into the interstellar 
medium (ISM) of the host galaxy and its environment, and is often invoked in 
theoretical and numerical simulations to reproduce the observed properties of 
galaxies \citep[e.g.][]{Croton06, Schawinski06}.    
%%%%
%%
At galactic scales, radio jets can regulate or prevent star formation i.e. provide 
the most commonly observed {\it negative} feedback by expelling or heating the gas.  
The radio jets can also trigger star formation i.e. provide {\it positive} feedback by shocking the gas to 
high densities thereby accelerating the gas cooling processes  
\citep[see e.g.][]{vanBreugel86, Chambers87, Daly90, Rejkuba02, Gaibler12}.    

%%%%%
While all these observational results confirm that the AGN feedback strongly influences the evolution 
of its host galaxy and the environment, the conditions prevailing in the host galaxy may also quench or 
trigger the AGN activity by altering the rate of accretion onto the central supermassive blackhole.  
One of the important questions in our understanding of the AGN is `What triggers the AGN activity?'.
%, and what are the timescales involved.
The galaxy-galaxy mergers represent one of the natural mechanisms to funnel large quantities of gas
to the central ($<$1\,pc) regions of galaxies and trigger the AGN.  
Several authors have found evidence for connection between the merger process and the AGN activity 
for  both radio-loud and 
radio-quiet AGNs \citep[e.g.][]{Combes09, Ellison11, Ramos-Almeida12, Villar-Martin12, 
Tadhunter12, Khabiboulline14}.  
However, there are also studies that find no relationship between the AGN activity and mergers 
\citep[e.g.][]{Cisternas11, Schmitt11}. 
In this context, it is of great interest to understand in detail the conditions that trigger AGN 
in galaxy mergers.    
%%%
It has been suggested that cold gas associated with the circumnuclear disk or torus may play a key 
role in fueling the AGN activity. 
Therefore, a natural way to test the connection between galaxy-galaxy mergers and the triggering of AGN 
activity will be to observe the properties of circumnuclear gas  in AGN hosts associated 
with mergers. 

%%%%
For radio-loud AGNs, an interesting way of probing the properties of circumnuclear atomic gas, i.e. its 
distribution and kinematics, is via \hi 21-cm absorption observations of the compact steep-spectrum (CSS) 
and gigahertz peaked-spectrum (GPS) 
sources\footnote{The CSS and GPS objects typically have sizes of $<$15 and 1\,kpc, respectively and believed 
to be young ($<$10$^5$\,yr) counterparts of the larger radio galaxies and QSOs \citep[][]{Odea98}. }.
There have been several \hi 21-cm absorption searches in samples of radio-loud AGNs involving 
these compact radio sources \citep[e.g.][]{Carilli98, VanGorkom89, Peck00, Morganti01, Vermeulen03,  
Gupta06, Chandola11, Allison12, Chandola13, Gereb15}. 
%%%
High \hi 21-cm absorption detection rate ($\sim$30-50\%), young age ($<$10$^5$\,yr) and the possibility of 
spatially resolved 21-cm absorption to trace the parsec scale distribution and kinematics of absorbing gas 
\citep[see Table~1 of][for a nice summary]{Araya10} makes these sources excellent candidates for understanding 
the triggering of AGN and its early stages of evolution.  

As CSS and GPS sources are still embedded in their host galaxy ISM, they are also ideally suited 
for understanding the impact of radio mode feedback in the form of radio jet-ISM interaction 
\citep[e.g.][]{Gelderman94, Axon00, Gupta05, Shih13}.      
In the \hi 21-cm absorption studies, the AGN feedback due to  CSS/GPS sources is revealed in the 
form of an excess of blue-shifted absorption lines with outflow velocities sometimes as high as 
 $\sim$ 1000\,\kms \citep{Vermeulen03, Morganti05, Gupta06, Chandola11, Mahony13, Morganti13, Gereb15}.
%%%
The incidences of redshifted or infalling gas clouds fuelling the AGN are rare \citep[e.g.][]{Araya10, Maccagni14}.

%%%%%%
In this paper, we report the detection and detailed analysis of the \hi 21-cm absorption associated with a 
merging pair of galaxies at $z\sim 0.123$ having an angular separation of 2.2 arcsec.  The absorption consists of 
a strong \hi 21-cm absorption component at the systemic redshift of the radio galaxy and 
two well-detached, infalling weaker components. 

This paper is structured as follows.  
In Section~2, we present details of optical and radio observations of the merging galaxy pair  
serendipitously discovered through our on-going search for 21-cm absorption from quasar-galaxy pairs 
\citep[][]{Gupta10, Gupta13}\footnote{Our sample of Quasar-Galaxy pairs at $z<0.3$ is 
selected by cross-correlating galaxies detected in SDSS with radio loud AGNs in FIRST.  There are 4
merging galaxy-galaxy pairs found. The 21-cm and optical observations of the remaining 3 sources are underway.}. 
In Section~3.1, we show that the optical emission line ratios of the galaxy associated with the 
radio source, SDSS J094221.98+062335.2 (refer to as J0942+0623 from now on), are similar to those of AGNs whereas the optical spectrum 
of the other galaxy is consistent with a normal star forming galaxy. 
Besides a strong H~{\sc i} 21-cm absorption component at the systemic velocity of the radio-loud AGN, we also detect two redshifted absorption components representing a gas falling into the 
AGN. 
Section~3.2 focusses on deriving the properties of these \hi absorption clouds from the arcsecond and 
milli-arcsecond scale spectroscopy.  
In Section~3.3, 
we discuss the relationship between the properties of \hi\ 21-cm absorption and mergers 
using a sample of compact radio sources with 21-cm absorption measurements.  
Finally in Section~4, we summarize our results.    
Throughout this paper we use the cosmological model with $\Omega_\Lambda$ = 0.73, $\Omega_M$ = 0.27 and 
H$_0$ = 71 km\,s$^{-1}$Mpc$^{-1}$. 

%%%%%%%%%%%%%%%%%%%%%%%%%%%%%%%%%%%%%%%%%%%%%%%%%%%%%%%%%%%%%%%%%%%%%%%%%%%%%%%%%%%%%%%%%%
\begin{figure*}
\centering
\psfig{figure=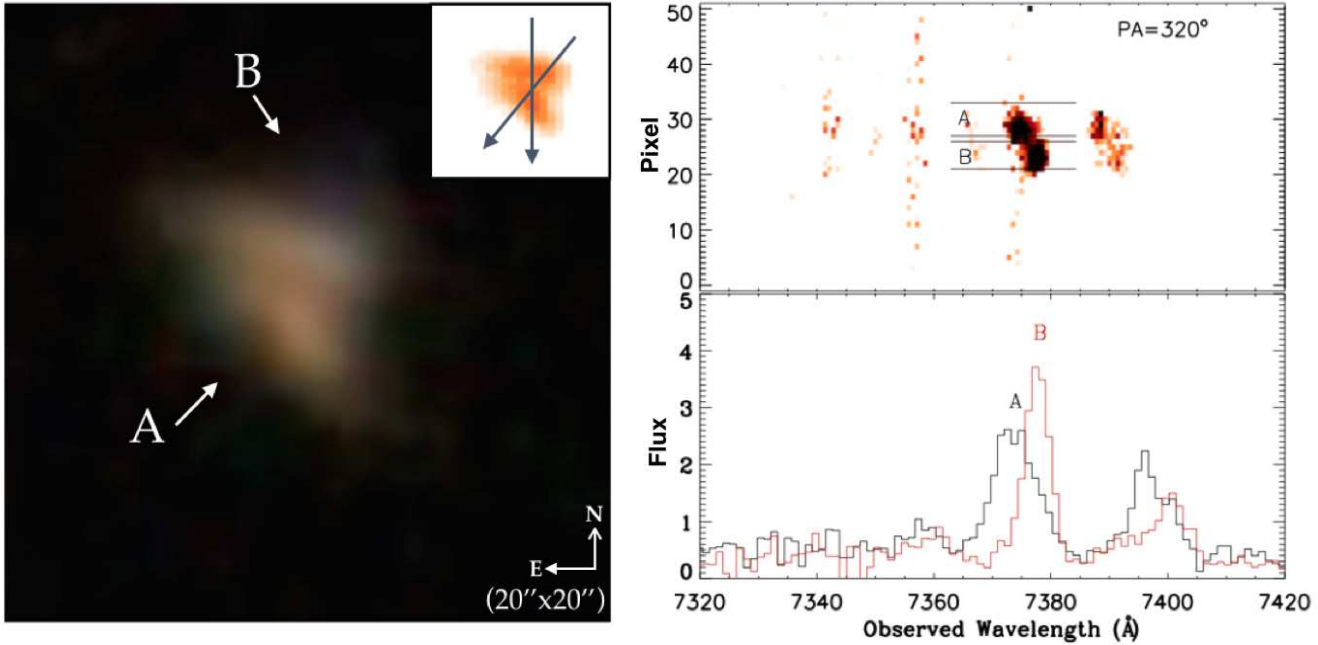,width=1.0\linewidth,height=0.5\linewidth,angle=0}
\caption{{\it Left}: SDSS colour composite image of the merging galaxy pair.  
{The image, 20$^{\prime\prime}$$\times$20$^{\prime\prime}$ in size, is centered at 
$\alpha_{\rm J2000}$ = 09\,h 42\,m 21.98\,s, $\delta_{\rm J2000}$ = 06\,d 23\,m 35.2\,s.} 
In the inset, we show the $\it r$-band image. The positions of two slits used for the 
IGO spectroscopic observations are marked. 
{\it Top-right}: The 2D spectrum covering the wavelength range of H$\alpha$ emission for the 
exposures with slit PA = 320$^\circ$. The two apertures used to extract the 1D spectra for the objects 
A and B are also marked. 
{\it Bottom-right}: One dimensional spectra corresponding to the two apertures showing the 
H$\alpha$ and [N~{\sc ii}] emission lines. 
}
\label{fig1}
\end{figure*} 
%%%%%%%%%%%%%%%%%%%%%%%%%%%%%%%%%%%%%%%%%%%%%%%%%%%%%%%%%%%%%%%%%%%%%%%%%%%%%%%%%%%%%%%%%%%%%%%%%%%%%

%%%%%%%%%%%%%%%%%%%%%%%%%%%%%%%%%
\begin{figure*}
\centering
\psfig{figure=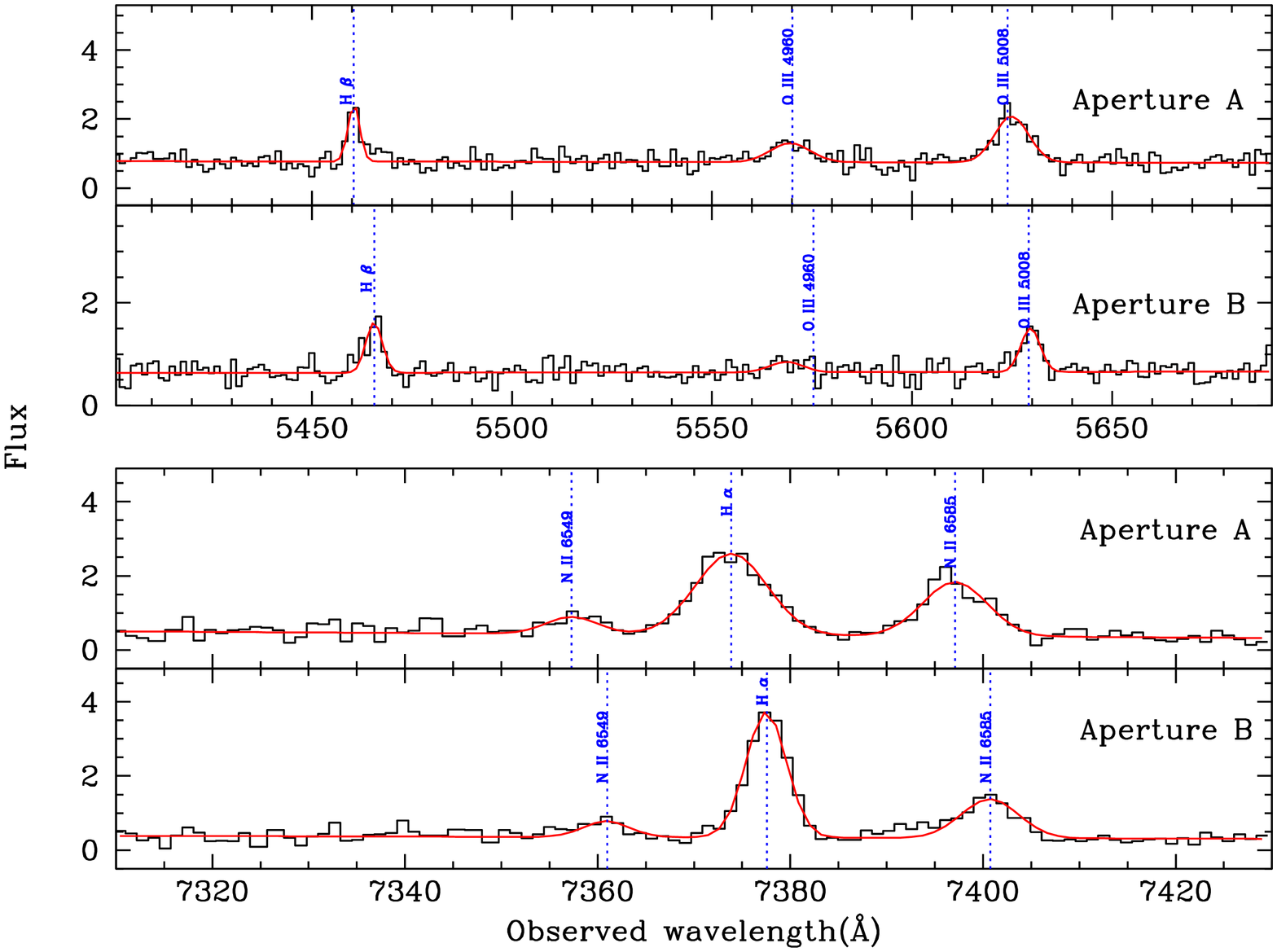,height=0.8\linewidth,angle=270}
\caption{IGO spectra of object A and B together with Gaussian fits to the 
\halpha, \hbeta, [N~{\sc ii}] and [O~{\sc iii}] lines. It is clear that the emission lines are broader in the case of
radio source (i.e. source A) which is at a lower redshift. 
} 
\label{fig2}
\end{figure*} 

%%%%%%%%%%%%%%%%%%%%%%%%%%%%%%%%%%%%%%%%%%%%%%%%%%%%%%%%%%%%%%%%%%%%%%%%%%%%%%%%%%%%%%%%%%%%%%%%%%%

%%%%%%%%%%%%%%%%%%%%%%%%%%%%%%%%%
\section{Observations and data analysis}

%%%%%%%%%%%%%%%%%%%%%%%%%%%%%%%%%
\subsection{Optical: IGO spectroscopy}
%%%%%%%%%%%%%%%%%%%%%%%%%%%%%%%%%

A single epoch optical spectrum of J094221.98+062335.2, observed on 27 February 2003, 
is available from the Sloan Digital Sky Survey (SDSS) archive. This spectrum covering 
a wavelength range of 
3800-9200\,\AA\ was obtained with a fibre of diameter 3$^{\prime\prime}$ centered 
at the position of the radio source and contains flux from both the galaxies.
Therefore, to determine the redshifts and the nature of both the objects, we 
performed the long-slit spectroscopy of J0942+0623 using the 
IUCAA Faint Object Spectrograph (IFOSC) mounted on the 2-m IUCAA Girawali Observatory (IGO) 
telescope. We used Grisms 7 and 8 of IFOSC\footnote{Details of the IGO/IFOSC grisms can be found at http://www.iucaa.
ernet.in/$\sim$itp/etc/ETC/help.html\#grism.} 
in combination with a long slit of 1.5$^{\prime\prime}$ width. These settings provide a wavelength coverage of 
3800$-$6840\,$\AA$  and 5800$-$8350\,$\AA$, and a spectral resolution of  
300 \kms and 240 \kms, respectively. The Grism 8 data includes three 45\,minutes exposures with the 
slit position angle, PA = 320$^\circ$ obtained on 25 February 2012 and three 45 minutes exposures 
with PA = 0$^\circ$ obtained on 26 April 2014. The Grism 7 data have one 45\,minutes 
exposure with the PA = 0$^\circ$ obtained on 27 January 2011.  The long slit orientations 
are shown in the {\it left} panel of Fig.~\ref{fig1}.
In the case of Grism 8, 
observations were carried out placing  the object at different locations along the slit 
to take care of the fringing
effects \citep[see][for details]{Vivek09}.

For these data, the cleaning of raw frames and 1D spectral extraction were carried out following 
standard procedures in {\sc iraf}{\footnote {IRAF is distributed by the National 
Optical Astronomy Observatories, which are operated by the Association of Universities 
for Research in Astronomy, Inc., under cooperative agreement with the National Science 
Foundation.}}. The two apertures that are shown overlaid on the 2-D spectra covering the H-$\alpha$ 
range 
in the {\it top-right} panel of Fig.~\ref{fig1} were chosen to extract  
1D spectra of the objects, A (the host galaxy of the radio source) and B. 
The 1D spectra covering the H$\alpha$ and [N~{\sc ii}] lines are shown in the {\it bottom-right} panel. 
The spectra covering H$\alpha$, H$\beta$, [N~{\sc ii}] and [O~{\sc iii}] lines overlaid
with Gaussian fits 
are shown in Fig~\ref{fig2}.
%%%
It is clear from these figures that the redshift of object A is less than that of B. 
We measure $z_A = 0.1232\pm0.0001$ and $z_B = 0.1239\pm0.0001$. 
The angular separation between the objects is 2.2 arcsec. 
This corresponds to a projected separation of 4.8\,kpc at the redshift of the radio source. 
The redshift difference measured corresponds to a velocity separation of
$\sim$185 \kms.The angular and velocity separations found here are consistent
with the definition of merging galaxies used by \citet{Ellison11} to study the
mergers in the SDSS database. In the following section, we study the line of
sight distribution of the cold H~{\sc i} gas in this merging system using H~{\sc i} 21-cm
absorption line as a tracer.

%%%%%%%%%%%%%%%%%%%%%%%%%%%%%%%%%
\subsection{Radio: WSRT and GMRT spectroscopy}
\label{sec:radobs}
%%%%%%%%%%%%%%%%%%%%%%%%%%%%%%%%%

%%%%%%%%%%%%%%%%%%%%%%%%%%%%%%%%%%%%%%%%%%%%%%%%%%%%%%%%%
\begin{figure*}
\includegraphics[height=14cm,angle=-90]{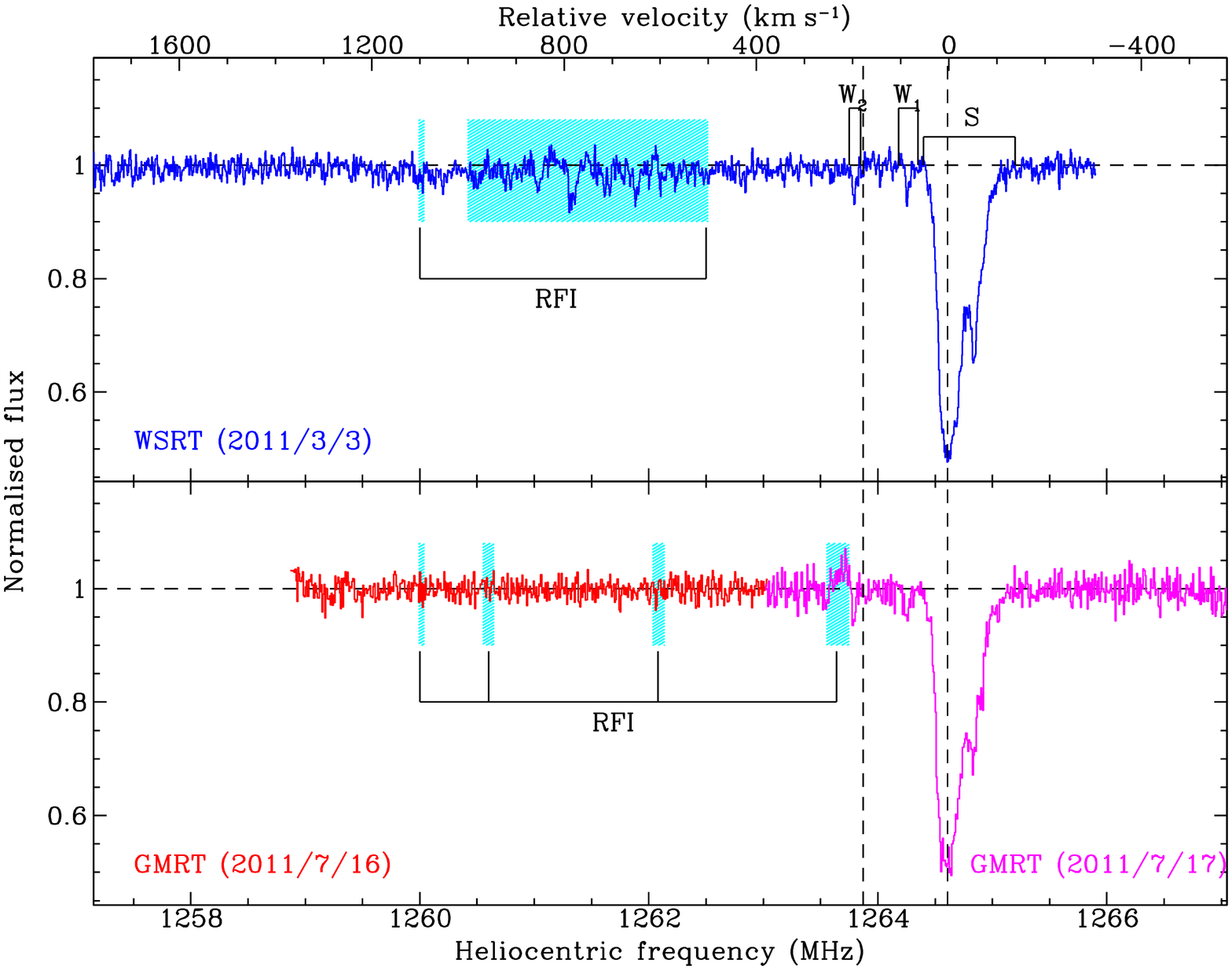} 
\caption{
The WSRT (top) and GMRT (bottom) spectra of the radio source i.e. object A.
The 21-cm absorption is clearly detected in a strong broad
component (marked as S) at the systemic velocity of the radio source and
two weak narrow absorption (marked as W$_2$ and W$_1$) with an
infalling velocity in the range $\sim$80-200\,\kms.  
The velocity scale shown in the top panel is defined with respect to 
the systemic redshift $z = 0.1232$ of the radio source A.  
The vertical dashed lines correspond to redshifts of objects A and B. 
%{\bf N: mark vertical lines and remove `blue wing'.} 
}
\label{21cm}
\end{figure*}
%%%%%%%%%%%%%%%%%%%%%%%%%%%%%%%%%%%%%%%%%%%%%%%%%%%%%%%%

We observed J0942+0623 with Westerbork Synthesis Radio Telescope (WSRT) 
on 3 March 2011 to search for the associated 21-cm absorption. 
A total bandwidth of 10\,MHz split into 2048 
spectral channels 
was used to acquire the data in the dual polarization mode.  
This corresponds to a total bandwidth of $\sim$2400\,\kms\ and a channel width of 
$\sim$1.2\,\kms\ prior to Hanning smoothing. 
The bright calibrators 3C\,286 and 3C\,147 were used for flux density and bandpass 
calibration.  The total on-source time, excluding calibration overheads, 
was $\sim$9.8\,hr.  The data were analysed using 
{\sc AIPS}\footnote{See http://www.aips.nrao.edu/index.shtml for 
more information about AIPS} following the standard procedures 
\citep{Gupta10}.  
%
%%%%%%
We detect 21-cm absorption spread over $\sim 300$ \kms\ in different components  
towards the radio source that is, compact in our WSRT maps having 
a resolution of 180$^{\prime\prime}\times$14$^{\prime\prime}$. 
The continuum normalised (stokes I) WSRT spectrum is shown in the {\it top} panel of
Fig.~\ref{21cm}. The measured continuum flux density is $\sim$113\,mJy.
The spectral resolution and the rms after applying Hanning smoothing
are 2.4\,\kms\ and 1.5\,mJy\,beam$^{-1}$\,channel$^{-1}$,  
respectively. The velocity scale, defined with respect to the \zem=0.1232 corresponding to 
the systemic velocity of the radio source (i.e. object A), is also plotted on the {\it top} axis of the panel. 

%%%%%%%
Redshifted 21-cm absorption line data are often affected by radio frequency interference (RFI).  
Our WSRT observations were no different in this regard.  The data were carefully examined to flag 
and remove the RFI.  However, certain frequency ranges were affected by the interference that 
was present on almost all the WSRT baselines for most of the times.  These frequency ranges are 
marked as shaded regions in the {\it top} panel of Fig.~\ref{21cm}, and affect the detectability 
of weak absorption features in the range: 1260-1263\,MHz.   

%%%%%%%
We observed the pair with Giant Metrewave Radio Telescope (GMRT) to 
confirm the weak \hi absorption features detected in the 
WSRT spectrum.  A bandwidth of 4\,MHz split into 512 frequency channels was used to cover 
the frequency ranges: 1259-1263\,MHz and 1263-1267\,MHz on 16 and 17 July 2011,  
respectively.  In both the observing runs, 3C\,147 was observed for flux and bandpass calibration.  
In addition, nearby compact radio source 0925+003 was observed every $\sim$45\,minutes for 
phase calibration.  In each observing run the total on-source time was $\sim$3\,hr.  
The data were analysed using {\sc AIPS} following standard procedures.  Relatively, the 
GMRT data were found to be much less affected by RFI.  The flux density of the radio source 
J0942+0623 was found to be $\sim$116\,mJy.  Within the relative uncertainties, this is 
consistent with the WSRT measurement.  The radio source is unresolved in the GMRT maps 
with a resolution of $\sim$3$^{\prime\prime}\times$2$^{\prime\prime}$.  The stokes-I continuum
normalised GMRT spectrum towards the radio source is shown in the {\it bottom} panel of 
Fig.~\ref{21cm}.  
The spectral resolution is $\sim$1.9\,\kms\ and the rms in the July 16 and 17 spectra are  
$\sim$1.7 and $\sim$2.3\,mJy\,beam$^{-1}\,$channel$^{-1}$, respectively.  
%%%%
Similar to the WSRT spectrum, the frequency ranges affected by RFI in most of the data are 
marked as shaded region in Fig.~\ref{21cm}.  
%%%%

%%%%%%%%
The velocity shift due to the heliocentric motion of Earth between the WSRT and GMRT observing
runs is 8\,\kms. This allows us to distinguish the {\it real} 21-cm absorption components from 
the features due to RFI. 
A strong absorption component, labelled as `S' in Fig.~\ref{21cm}, and a weaker 
component, labelled as `W$_1$', are detected in both the WSRT and GMRT 
spectrum at the observing frequency consistent with the heliocentric shift.  
Another component W$_2$, although very close to the RFI `spike' 
in the GMRT spectrum, is consistently reproduced in both the spectra.  
None of the other weaker features in the frequency range 1260-1263\,MHz seen in the
WSRT spectrum are reproduced in the GMRT spectrum and we consider these to be due to RFI.

The properties of absorption components S, W$_1$ and W$_2$ i.e. the absorption width, integrated 
21-cm optical depth ($\int\tau dv$) and \hi column density ($N$(H~{\sc i})) are summarized in Table~\ref{tab21cm}.  
Note that the inferred large $N$(H~{\sc i})$\sim$10$^{22}$\,cm$^{-2}$ is extremely rare 
in the samples of QSO sight lines tracing the ISM/halos of {\it normal} intervening galaxies 
via 21-cm, Mg~{\sc ii} or damped Lyman-$\alpha$ (DLA) absorption \citep{Gupta10, Gupta12, Srianand12dla, Noterdaeme12dla}.    
However, such $N$(H~{\sc i}) are more commonly associated with Gamma-ray bursts \citep[e.g.][]{Jakobsson06}, 
and galaxies at an impact parameter of a few kpc \citep[][]{Noterdaeme14} i.e. the regions of high density 
close to the center of galaxy.
%%%
Amongst the known 21-cm absorbers associated with  AGNs such column densities  (i.e. $\sim$10$^{22}$\,cm$^{-2}$)
have been seen in a few cases i.e. J0111+3906 (also known as OC314), SDSS J150805.7+342323.3, UGC\,6081,  UGC\,8387 and Mrk\,273 
\citep{Carilli98, Chandola11, Darling11, Gereb15}.  
It is clear from SDSS images that the host galaxies associated with J1508+3423 at $z=0.046$, 
UGC\,6081 at $z=0.036$, UGC\,8387 at $z=0.023$ and Mrk\,273 at $z=0.037$ are undergoing a 
major merger.
In the case of higher redshift ($z=0.6685$) source 
J0111+3906, the host galaxy shows asymmetric morphology \citep[][]{Stanghellini98}, 
possibly a signature of an ongoing 
or a recent merger.  
As galaxy mergers are expected to trigger the inflow of gas to the central regions of the galaxies,  
the large columns of \hi gas associated with J0942+0623 and these {five} cases from the 
literature are caused by the ongoing mergers.

%%%%%%%%%%%%%%%%%%%%%%%%%%
\begin{table}
\caption{Properties of the 21-cm absorption components as measured from the WSRT spectrum}
\begin{tabular}{cccccc}
\hline
\hline
Component  & Width$^\dagger$  &  $\int \tau(v) dv$ & $N$(H~{\sc i})(100/T$_s$)($f_c$/1.0)$^\ddag$ \\
           & (\kms) & (\kms)             &  (10$^{20}$\,cm$^{-2}$)                   \\
\hline
S             &   98   & 49.9$\pm$0.4       &     91.0$\pm$1.0                       \\
W$_1$         &   21   & 0.54$\pm$0.09      &     0.99$\pm$0.15                       \\  
W$_2$         &   14   & 0.63$\pm$0.07      &     1.15$\pm$0.15                       \\
\hline
\end{tabular}
\begin{flushleft}
$^\dagger$  Velocity width corresponding to 90\% of the total $\int\tau$dv.\\
$^\ddagger$ T$_s$ = spin temperature; $f_c$ = covering factor. 
\end{flushleft}
\label{tab21cm}
\end{table}
%%%%%%%%%%%%%%%%%%%%%%%%%%%%%

%%%%%%%%%%%%%%%%%%%%%%%%%%%%%%%%%
\subsection{Radio: VLBA imaging and spectroscopy}
%%%%%%%%%%%%%%%%%%%%%%%%%%%%%%%%%
%
%
%%%%%%%%%%%%%%%%%%%%%%%%%%%%%%%%%%%%%%%%%%%%%%%%%%%%%%%%%
\begin{figure}
\vbox{
\includegraphics[height=8.5cm,angle=-00]{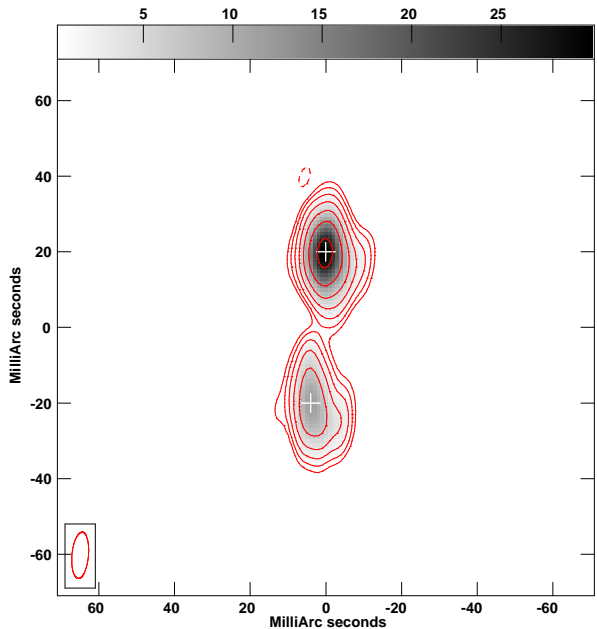} 
}
\caption{
VLBA image of J0942+0623.  
The rms in the image is 90\,$\mu$Jy\,beam$^{-1}$. The contour levels are 
4.0$\times$10$^{-4}$$\times$($-1$, 1, 2, 4, 8, ...)\,Jy\,beam$^{-1}$. 
The gray scale range is: 0.4$-$30\,mJy\,beam$^{-1}$.  
The restoring beam (see text for details) is shown as an ellipse at the bottom left corner. 
The image centre is at RA = 09$^{\rm h}$\,42$^{\rm m}$\,21$^{\rm s}$.9832 and 
Dec. = +06$^\circ$\,23$^\prime$\,35$^{\prime\prime}$.1150. The northern and southern 
peaks are marked with '+'. They are separated by $\sim$40\,mas i.e. 89\,pc. 
}
\label{vlbamap}
\end{figure}
%%%%%%%%%%%%%%%%%%%%%%%%%%%%%%%%%%%%%%%%%%%%%%%%%%%%%%%%
%%%%
The source J0942+0623 was observed with Very Long Baseline Array (VLBA) on 2012 June 15, 17 and 22.  
The goal of these observations was to perform the milliarcsecond resolution 1264\,MHz continuum
imaging 
and the redshifted \hi 21-cm absorption spectroscopy.  
The total observing time was 10\,hr.  
The on-source time, excluding the telescope set-up time and calibration overheads, 
was $\sim$6 hr. One 8\,MHz-baseband 
channel pair covering the frequency range 1260.8-1268.8\,MHz was used with right- and 
left-hand circular polarizations sampled at 2 bits.  
The data were correlated with 4000 spectral channels providing a spectral resolution 
of 2\,kHz i.e. $\sim$0.47\,\kms.  The correlator integration time was 2\,s.  
We analysed the data following standard procedures in {\sc aips} \citep[see for example][]{Momjian02, Gupta12}.  
%%%
The continuum image of the radio source obtained by averaging absorption-free frequency 
channels is shown in Fig.~\ref{vlbamap}.
%%%%
The rms in the image is 90\,$\mu$Jy\,beam$^{-1}$, and the spatial resolution i.e. 
the restoring beam is 0.01230$^{\prime\prime}$$\times$0.00436$^{\prime\prime}$ with 
PA = $-$5.1$^\circ$. 

%%%%% 
In the VLBA image, we recover $\sim$ 88\% of the total flux density detected in the GMRT and
WSRT images. Most of the milliarcsecond scale flux is contained in the two components which 
are marked with `+' in the Fig.~\ref{vlbamap}.  Hereafter, we refer to these as the northern 
and southern blobs, or VLBA-N and VLBA-S, respectively.  
The separation between the peaks of VLBA-N and VLBA-S is $\sim$40\,mas i.e. 89\,pc at $z_g=$0.1232.   
The overall radio morphology is similar to the Compact Symmetric Objects (CSOs) believed to 
be young ($<$10$^5$\, yr old) radio sources \citep[e.g.,][]{Conway02}. 
The peak and total flux densities of VLBA-N and VLBA-S are provided in Table~\ref{vlba_tab1}.  
Both the blobs are spatially resolved and each can be fitted with two Gaussian components.  
We find that in each case more than 70\%  of the flux in the
blob is contained in a Gaussian component coinciding with the peak with FWHM $<$ 10 mas.  
%%%%%
%
%%%%%%%%%%%%%%%%%%%%%%%%%
\begin{table}
\caption{Results of VLBA observations}
\begin{tabular}{ccccc}
\hline
\hline
Source       & Peak & Integrated &  $\int\tau dv$$^\dag$\\
component    & flux density & flux density &        \\
      & (mJy\,beam$^{-1}$) & (mJy) & (km\,s$^{-1}$) \\
\hline
VLBA-N & 31.0 & 68.9 &  33$\pm$3 \\
VLBA-S & 11.9 & 31.4 &  118$\pm$20 \\
\hline
\end{tabular}
\begin{flushleft}
$^\dagger$  Towards VLBA-N and VLBA-S peaks.
\end{flushleft}
\label{vlba_tab1}
\end{table}
%%%%%%%%%%%%%%%%%%%%%%%%%%

%%%%%%%
%
In the top panel of Fig.~\ref{vlbaspec}, we plot VLBA spectrum extracted towards the radio continuum 
detected in Fig.~\ref{vlbamap}.  The WSRT and integrated VLBA spectra match well within the uncertainties.  
This suggests that all the 21-cm absorption (except for W$_1$ and W$_2$ which are too weak to be detected 
in the VLBA spectrum) detected in the WSRT/GMRT data are recovered in the VLBA data, and no absorption 
is present towards the radio flux resolved out in the VLBA data.  The 21-cm absorption spectra towards the
 VLBA-N and VLBA-S peaks are shown in the bottom two panels of Fig.~\ref{vlbaspec}.
%
%
%
%%%
As the optical depth sensitivity falls off rapidly beyond the continuum peaks, 
instead of making spatially resolved optical depth maps, we will make detailed comparison of 
the spectra towards the VLBA-N and VLBA-S peaks with our GMRT and WSRT spectra.  These 
results are presented in the next Section.

%%%%%%%%%%%%%%%%%%%%%%%%%%%%%%%%%%%%%%%%%%%
\section{Results and Discussion}
%%%%%%%%%%%%%%%%%%%%%%%%%%%%%%%%%%%%%%%%%%%

In this section we present a detailed analysis of the optical and radio spectra and
discuss the physical state of the merging galaxies and the \hi gas producing the 21-cm 
absorption.

%%%%%%%%%%%%%%%%%%%%%%%%%%%%%%%%%%%%%%%%%%%
\subsection{The merging galaxy pair: SDSS J0942+0623}
%%%%%%%%%%%%%%%%%%%%%%%%%%%%%%%%%%%%%%%%%%%

We use Gaussian fits to the emission lines to measure the line parameters towards
the objects A and B. For the object A, we measure the FWHM of the H$\alpha$ line to be 350 \kms.
We find log~[{[O~{\sc iii}]$\lambda$5007/${\rm H\beta}$}]$\sim$0.4 and log~[[N~{\sc ii}]$\lambda6583$/H~$\alpha$]$\sim-0.2$. These values are consistent with the source A being a narrow line AGN
\citep[see Fig~14.2 of][]{Osterbrock06}. As per the definition of \citet{Kewley06}, the 
source A can be classified as a composite H~{\sc ii}-AGN type galaxy.
In comparison, the emission lines associated with the object B are typically narrow and barely resolved in the IGO spectrum. We find
 log~[{[O~{\sc iii}]$\lambda$5007/${\rm H\beta}$}]$\sim$ 0.0 and log~[[N~{\sc ii}]$\lambda6583$/H~$\alpha$] $\sim-0.4$. These values are consistent with what one expects from a
H~{\sc ii} region like object. Thus there are no signatures of any strong AGN activity
in object B. Using the inferred H$\alpha$ luminosity we estimate the dust uncorrected
starformation rate \citep[using equations in][]{Argence09} of $\sim$0.4 M$_\odot$ yr$^{-1}$. This suggests that there is no strong star-bursting activities in the 
central regions of object B.
Thus, the objects A and B form an interacting pair of galaxies, with a separation of 4.8\, kpc, in which the former is a radio-loud narrow emission line 
AGN whereas the latter is a normal star-forming galaxy.

%%%
It is not unusual for an AGN to be associated with a host galaxy undergoing a merger.  
Several authors have found evidence for mergers to be associated with both radio-loud and 
radio-quiet AGNs \citep[e.g.][]{Combes09, Ellison11, Ramos-Almeida12, Villar-Martin12, 
Tadhunter12, Khabiboulline14}.   
In a few merging pairs, double AGNs have also been found \citep[e.g.][]{Evans08, Tadhunter12}. 
%%%
Such systems are though extremely rare. In general, there could be several reasons for the lack of 
simultaneous AGN activities in both the merging galaxies like (i) intermittent
supply of gas to the central regions or (ii) suppression of gas infall due to
negative feedback caused by the nuclear starburst activity.
 In the case of object B, the lack of AGN activity is possibily due to 
the lack of cold gas supply to the central region. 

%%%%%%%%%%%%%%%%%%%%%%%%%%%%%%%%%%%%%%%%%%%
\subsection{Properties of the absorbing clouds}
%%%%%%%%%%%%%%%%%%%%%%%%%%%%%%%%%%%%%%%%%%%

In this section, we first discuss the properties of component `S'  
inferred from the arcsecond and milliarcsecond scale 21-cm absorption spectra. 
Following this we present the properties of well-detached narrow
components W$_1$ and W$_2$ that are infalling with respect to the systemic
redshift of the radio source. These features are detected only in the
more sensitive GMRT and WSRT spectra.
%

%%%%%%%%%%%%%%%%%%%%%%%%%%%%%%%%%%%%%%%%%%%
\subsubsection{The 21-cm absorption component: S}
\label{sec:comps}
%%%%%%%%%%%%%%%%%%%%%%%%%%%%%%%%%%%%%%%%%%%

%%%%%
The 21-cm absorption component `S' extends over the frequency range: 1264.4-1265.2\,MHz.  
The absorption profile exhibits substructure indicating multiple absorption components.  
 As is evident in Fig~\ref{21cm}, the absorption profile is asymmetric with respect
to the systemic redshift of the radio source with the blue wing extending upto $-$120\,\kms
and the red wing upto +50\,\kms.
The 90\% of the total optical depth is contained within $\sim$98\,\kms\ and the optical depth 
weighted centroid of the profile is almost (within $\sim$20\,\kms) coincident with the 
systemic redshift of the radio source.   

%%%
Recently, \citet[][]{Gereb15} used 21-cm absorption line shapes 
and systemic velocities for a sample of 32 absorbers to study
the origin of the absorbing gas.  They suggested that while the narrowest 
lines with ``busy function'' FWHM$<$100\,\kms\ detected close to the 
systemic velocities are likely to be 
produced by \hi disks, the absorption with intermediate and large (FWHM$>$200~\kms) widths originate 
from \hi disks with complex gas kinematics. 
%and/or gas interacting with the radio source.
%%
Among the 21-cm detections in their sample, three sources (UGC\,05101, UGC\,8387 and Mrk\,273) 
are associated with mergers, and all three exhibit extremely broad absorption lines spreading over  $\sim$500-1000\,\kms. 
Also, unlike J0942+0623 which is a CSO with an extent of 89\,pc, the radio continuum 
associated with these mergers have complex morphology with much 
larger extents \citep[][]{Condon90, Crawford96, Carilli00} suggesting that the absorbing 
gas is originating from gas with complex kinematics due to ongoing merger and interaction 
with the radio source. 
The properties of component `S' are akin to the class of \hi absorbers with intermediate widths 
detected close to the systemic velocity.  This suggests that the absorption originates from a circumnuclear disk having simple kinematics. However, 
the narrower \hi absorption line in our case can be attributed
to the compactness of the J0942+0623 which may be
tracing only a small portion of \hi gas velocity distribution.

\begin{figure}
\vbox{
\includegraphics[height=8.5cm,angle=-00]{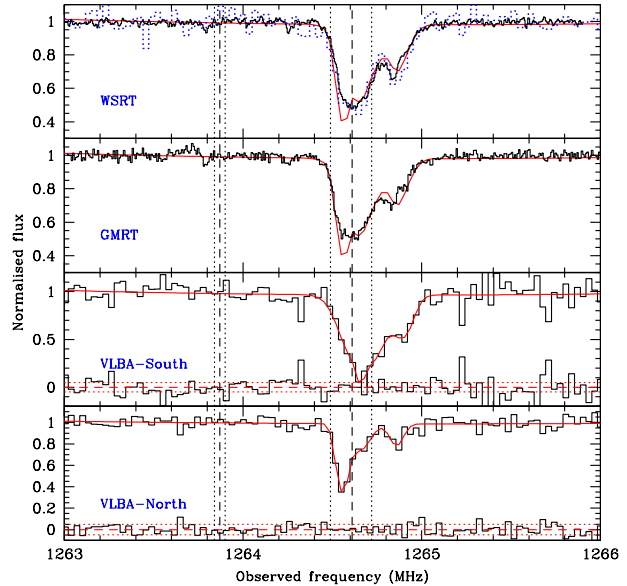} 
}
\caption{
VLBA spectrum towards the northern (VLBA-N) and southern (VLBA-S) peaks marked in 
Fig.~\ref{vlbamap} are shown in bottom two panels. Multiple Gaussian fit to the
21-cm absorption is also overlaid. In the bottom of each of these panel residual
flux resulting from the fits are compared to the rms flux measured in the line
free regions. Top two panels show the GMRT and WSRT spectra. In these, we 
overlay the total VLBA spectrum {\it predicted} using spectra towards VLBA-N and VLBA-S 
peaks (see text for details). The vertical dashed and dotted lines mark the emission redshifts and associated errors respectively
of object A and B.  The integrated VLBA spectrum extracted towards the radio continuum detected in Fig. 4 is shown as dotted line in the top panel.
}
\label{vlbaspec}
\end{figure}

%%%%%%
To understand this further, we compare the WSRT/GMRT spectra with the 
VLBA spectra towards the peaks of VLBA-N and VLBA-S in Fig.~\ref{vlbaspec}.  
The differences in the optical depth are clearly seen towards these two sight lines separated 
by 89\,pc (see bottom two panels of Fig.~\ref{vlbaspec}). 
The total integrated optical depth towards VLBA-S peak is higher by a factor of 
$\sim$3.5 compared to that measured towards VLBA-N peak (see also Table~\ref{vlba_tab1}).  

%%%
The top two panels in Fig.~\ref{vlbaspec} show the GMRT and WSRT spectra. First we check whether
the individual absorbing gas detected towards the VLBA peaks cover all the 
radio emission from the corresponding VLBA component. For this we assume
that the total flux seen in the WSRT/GMRT image is the sum of the fluxes of
the two VLBA blobs. This allow us to predict the expected absorption
profile in the WSRT/GMRT spectrum i.e.,
\begin{equation}
I_\nu = 0.314~exp (-\tau_S(\nu)) + 0.686~exp(-\tau_N(\nu)).
\label{eq1}
\end{equation}
Here, I$_\nu$ is the normalised flux measured with WSRT/GMRT and $\tau_N$ and
$\tau_S$ are the optical depth measured towards VLBA-N and VLBA-S respectively.
The values 0.314 and 0.686 are the fractional radio flux contained in the VLBA-S and VLBA-N
respectively.

The predicted absorption profile (solid smooth profile) using Eq.~\ref{eq1}, is overlaid on the 
WSRT/GMRT spectra in the top two panels. It is clear that apart from the location where
the peak absorption occurs towards VLBA-N, the predicted profile matches well 
with the observed one. This implies that the projected extent of the gas 
responsible for most of the absorption is at least as big as the individual continuum
emitting regions in the VLBA images (i.e $>20$pc). However, the above noted
inconsistency over a narrow velocity range could be due to the presence of absorption
by compact gas with dimensions smaller
than 20\,pc. 

%%%%%%%%%%%%%%%%%%%%%%%%%%%%%%%%%%%%%%%%%%%%%%%%%%%%%%%%%
%%%%%%%%%%%%%%%%%%%%%%%%%%%%%%%%%%%%%%%%%%%%%%%%%%%%%%%%

%%%%%%%%%%%%%%%%%%%%%%%%%%%%%%%%%%%%%%%%%%%%%%%%%%%%%%%%%%
%\begin{table}
%\begin{center}
%\caption{Results of multiple gaussian fits to the 21-cm absorption 
%detected towards the VLBA-N and VLBA-S peaks}
%\begin{tabular}{cccccc}
%\hline
%\hline
%ID & \zabs & $\tau_p$ & FWHM    & T$_{kin}^{max}$ & $\int \tau dv$\\
%   &   &          & (\kms)  & (K)           & \kms\\ 
%\hline
%\multicolumn{6}{c}{Fits for the profile towards VLBA-N }\\
%N1   & 0.123245 & 0.994 & 16.75 & 6130  & 17.8\\
%N2   & 0.123158 & 0.292 & 30.21 & 19960 & 19.4\\
%N3   & 0.122979 & 0.245 & 20.58 & 9260  &  5.4\\
%\multicolumn{6}{c}{Fits for the profile towards VLBA-S }\\
%S1   & 0.123163 & 1.73 & 9.18 & 1840 & 17.0\\
%S2   & 0.123140 & 1.58 & 50.47 & 55680 & 85.2\\
%S3   & 0.122941 & 0.56 & 27.81 & 16920 & 16.6\\
%\hline
%\end{tabular}
%\label{vlba_tab2}
%\end{center}
%\end{table}
%%%%%%%%%%%%%%%%%%%%%%%%%%%%%%%%%%%%%%%%%%%%%%%%%%%%%%%%%

%%%%%%%%%%%%%%%%%%%%%%%%%%%%%%%%%%%%%%%%%%%%%%%%%%%%%%%%%%
\begin{table}
\begin{center}
\caption{Results of multiple gaussian fits to the 21-cm absorption 
detected towards the VLBA-N and VLBA-S peaks}
\begin{tabular}{cccccc}
\hline
\hline
ID & \zabs & $\tau_p$ & FWHM    & T$_{kin}^{max}$ & $\int \tau dv$\\
   &   &          & (\kms)  & (K)           & \kms\\ 
\hline
\multicolumn{6}{c}{Fits for the profile towards VLBA-N }\\
N$_1$   & 0.123245(1) & 0.98(0.10) & 17.04(1.27) & 6350  & 17.8(0.3)\\
N$_2$   & 0.123159(2) & 0.29(0.03) & 30.22(9.02) & 21315 & 19.6(0.5)\\
N$_3$   & 0.122975(5) & 0.24(0.04) & 19.70(3.59) & 8490  &  5.0(0.2)\\
\multicolumn{6}{c}{Fits for the profile towards VLBA-S }\\
S$_1$   & 0.123154(1) & 1.72(0.16) & 9.36(1.33)  & 1916  & 17.2(0.5)\\
S$_2$   & 0.123139(2) & 1.56(0.05) & 50.36(1.70) & 55450 & 83.6(0.6)\\
S$_3$   & 0.122940(4) & 0.56(0.03) & 26.38(2.13) & 15210 & 15.4(0.2)\\
\hline
\end{tabular}
\label{vlba_tab2}
\end{center}
\end{table}
%%%%%%%%%%%%%%%%%%%%%%%%%%%%%%%%%%%%%%%%%%%%%%%%%%%%%%%%%

%%%%%%%%%%%%%%%%%%%%%%%%%%%%%%%%%%%%%%%%%%%%%%%%%%%%%%%%%
\begin{figure}
\vbox{
\includegraphics[height=8.5cm,angle=-00]{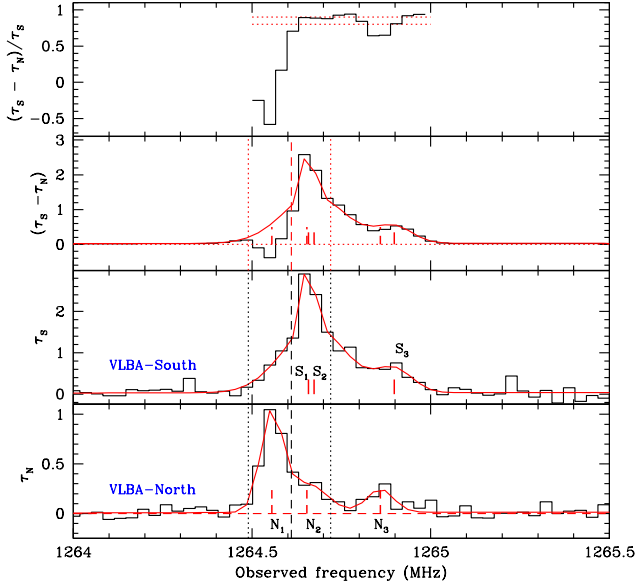} 
}
\caption{Measured H~{\sc i} 21-cm absorption optical depth towards the two 
VLBA components are shown in the bottom two panels. Optical depth
difference is shown in the second panel from the top. The top panel
shows the fractional optical depth excess towards the southern component.
Central frequency of individual Gaussian components are marked. 
The vertical dotted dashed lines are as in Fig~\ref{vlbaspec}.
In the top panel the dotted horizontal lines  give the mean fractional
variation in $\tau$ over the frequency range 1264.6 MHz and 1265 MHz.
}
\label{vlbaspec1}
\end{figure}
%%%%%%%%%%%%%%%%%%%%%%%%%%%%%%%%%%%%%%%%%%%%%%%%%%%%%%%%

%%%%%%
Next we decompose the 21-cm H~{\sc i} absorption profiles towards VLBA-N and VLBA-S peaks using 
multiple Gaussians.  
We use these to examine if the properties of \hi gas can be explained by a simple disk model.
%  the optical depth gradients in the H~{\sc i} gas.
The fits are summarized in Table~\ref{vlba_tab2}, and are plotted in the bottom two 
panels of Fig.~\ref{vlbaspec1}.
Both the profiles are well fitted by three components.  We label these components as N$_1$ to N$_3$ 
and S$_1$ to S$_3$ for VLBA-N and VLBA-S, respectively.  

It is apparent that a Gaussian component at $z=0.123245$ corresponding to N$_1$ 
is not required to fit the profile towards VLBA-S. This probably corresponds 
to the narrow component over-predicted by the VLBA profile noted earlier and arises from a 
distinct cloud that at best covers only the VLBA-N.  
Similarly the component S$_1$ is seen only towards VLBA-S.  For an assumed spin temperature 
of 100\,K, the log\,$N$(H~{\sc i}) in both the cases is $\sim$21.5.  If we assume the gas 
clouds to be spherical and covering only VLBA-N or VLBA-S then this will correspond to 
a gas density of $n_H$$\sim$50\,cm$^{-3}$.  
The density will be higher if T$_S$$\gg$100\,K as expected in the vicinity of a strong 
radio continuum source.   
For typical jet advancement speeds of $\sim$0.1c observed in the case of compact radio 
sources, the ambient medium with clouds of such densities can confine the 
radio source \citep{Taylor00, Bicknell03, Gugliucci05, Jeyakumar05}.  
However, kinematically both N$_1$ and S$_1$ are consistent with the systemic velocity of the host 
galaxy within $\sim$10\,\kms.  This implies that these clouds are most likely associated with the 
circumnuclear disk or torus rather than the outflowing gas clouds interacting with the radio source.   
%%

%%%%%%
%%%
The other two Gaussian components fitted to the spectrum towards VLBA-N and VLBA-S 
comprise $\sim$75\% of the total optical depth.   
These form a more coherent structure in opacity and kinematics that is consistent with a 
circumnuclear disk/torus like structure.  
The components N$_2$ and N$_3$ are within 10\,\kms\ to the components S$_2$ and S$_3$, respectively. 
This implies that there is no strong velocity gradient in the absorbing gas
along the jet-axis i.e. the north-south direction.  

In top two panels of Fig.~\ref{vlbaspec1}, we plot the fractional optical depth, 
($\frac{\tau_S - \tau_N}{\tau_S}$) and the difference optical depth, ($\tau_S - \tau_N$) 
spectra. In the difference plot, we overlay the multiple Gaussian fit towards VLBA-S 
scaled by 0.85, the mean of ($\frac{\tau_S - \tau_N}{\tau_S}$) in the velocity range 
corresponding to S$_2$ and S$_3$.  
Clearly the scaled $\tau_S$ fits match well with the difference spectra (i.e $\tau_S - \tau_N$).  
This and the kinematical coincidence 
between N2 (N3) and S2 (S3) imply that these components cover both VLBA-N and VLBA-S 
albeit with an optical depth gradient from VLBA-N to VLBA-S i.e. over 89\,pc.       
Following the circumnuclear disk model illustrated in the Fig.~5 of \citet{Orienti06}, 
the opacity gradient across the radio source can be explained by a combination of 
torus half-opening angle and, most importantly, the orientation of the line of sight 
such that the sight line towards VLBA-S traverses much longer path length through the 
disk \citep[see also][]{Gupta06uni, Curran10assoc}.

Thus, the VLBA spectroscopic observations are consistent with the absorption
component `S' originating from a clumpy circumnuclear disk
\citep[as seen by][in the case of NGC~7674]{Momjian03}.

%%%%%%%%%%%%%%%%%%%%%%%%%%%%%%%%%%%%%%%%%%%
\subsubsection{The \hi 21-cm absorption components: W$_1$ and W$_2$}
%%%%%%%%%%%%%%%%%%%%%%%%%%%%%%%%%%%%%%%%%%%

We detect 21-cm absorption component W$_1$ and W$_2$ in the WSRT and GMRT spectrum.  
With respect to the radio source, these components have  infall velocities of 
$\sim$+85 and $\sim$+195\,\kms\, respectively. 
%%%
The widths corresponding to 90\% of the total optical depth and the integrated optical 
depths of these components are summarized in Table~\ref{tab21cm}.  The narrow
velocity spread seen is consistent with the infalling gas being clumpy
rather than a smooth infalling distribution of gas where one expects to detect broad absorption lines.
Typically, hotspot separation speeds of $\sim$0.1-0.5c have been observed for CSOs \citep[e.g.][]{Taylor00}.   
Adopting a value of 0.1c for J0942+0623 we obtain the kinematic age of $\sim$1500 yrs. 
This recent AGN activity may have been triggered by the infalling gas 
whose signatures are still seen in the form of W$_1$ and W$_2$.

%%%%
The component W$_1$ at \zabs = 0.1235 is well represented by a single Gaussian component of FWHM of 
6.4$\pm$1.0 \kms. 
This places a constraint on the kinetic temperature to be less than 900~K and correspondingly 
log~$N$(H~{\sc i})$\le$20.95\footnote{Here we assume that the line width provides the maximum allowed temperature and spin temperature is equal to the gas kinetic temperature.}.
%%%
Similarly, the component W$_2$ at \zabs = 0.12393 can also be fitted well with  
a single Gaussian component of FWHM of 11.5$\pm$1.5~\kms. This places a constraint on the 
kinetic temperature to be less than 2900\,K and correspondingly log~$N$(H~{\sc i})$\le 21.46$.
%%%
From Table~\ref{tab21cm} it is clear that for any realistic 
spin-temperature the inferred $N$(H~{\sc i}) will place the absorbing gas in the
category of sub-DLAs or DLAs.
The upper limit on the kinetic temperature derived for both these components are 
consistent with the infalling gas being cooler than what is expected  (i.e. $10^4$ K)
for an intergalactic H~{\sc i} gas in photoionization  equilibrium with the UV background
radiation. 
%

%%%
{ However, instead of being part of the infalling gas feeding the AGN,  components W$_1$ and W$_2$ 
could simply be from (1) an intervening tidally disrupted gas or 
(2) the gas associated with objects A and B, respectively.  
This second possibility can be understood as following. 
As shown in Fig.~\ref{21cm}, the blue wing of component `S' extends upto +120\,\kms.  
The infall velocity (+85\,\kms) of W$_1$ is well within this range and it is possible that 
W$_1$ also originates from the same circumnuclear disk as `S'.  
Similarly, the component W$_2$ appears blueshifted with respect to object B by $\sim$10\,\kms.  
But this is within the errors ($\sim$30\,\kms) on the redshift of object B 
implying that W$_2$ may just be associated with object B.
}

%%%%%%%%%%%%%%%%%%%%%%%%%%%%%%%%%%%%%%%%%%%
\subsection{Mergers and triggering of radio-loud AGN}
\label{sec:merg}
%%%%%%%%%%%%%%%%%%%%%%%%%%%%%%%%%%%%%%%%%%%

%%%%%%
It has already been established through various 21-cm absorption line surveys 
mentioned in Section~1 that detection rate  of intrinsic 21-cm absorption amongst the
compact radio sources is significantly higher ($\sim$50\%) than in the 
extended sources (10-20\%). 
This has been interpreted as the evidence of gas rich environment in these younger 
sources triggering the AGN activity. 
%%%%
In this context, it is interesting of ask `what fraction
of merging galaxies associated with compact radio
sources are detected in \hi 21-cm absorption detections?'.

%%%%
This question can be addressed by considering the \hi absorption study by \citet[][]{Chandola11}. 
The study is based on 18 CSS and GPS sources drawn from the CORALZ sample \citep[Compact Radio Sources at 
Low Redshift;][]{Snellen04}. As all the sources in this study are at low-$z$ ($z<0.15$) and 16/18  
are also covered in SDSS, it is possible to determine the merger fraction for these sources by visual 
inspection.  
We find that three CSS/GPS sources, namely, J1409+5216, J1508+3423 and J1718+5441 are associated with 
galaxy-galaxy mergers and exhibit morphological distortions and tidal features typically showed 
by such systems.  
All three are detected in 21-cm absorption with \hi column densities in the range: 
10$^{21-22}$\,cm$^{-2}$, which is at the higher end of column densities detected in their sample. 
This is also true for three mergers from the sample of \citet[][]{Gereb15} discussed in 
Section~\ref{sec:comps}. 
%%%%
All this further reinforces the point already mentioned in Section~\ref{sec:radobs} 
that extremely high \hi column densities ($\sim$10$^{22}$\,cm$^{-2}$), as seen in the case of 
J0942+0623, appear to be associated with galaxies undergoing major mergers. 
It is important to note that, in general at $N$(\hi)$<$10$^{22}$ cm$^{-2}$, one does see a mix 
of mergers and non-mergers, and the fraction of non-merging systems increases as the $N$(\hi) cut-off is 
lowered.    The major limitation at this point in quantifying this and drawing general 
conclusions is the unavailability of good quality optical imaging and spectroscopic data 
to reliably identify galaxy companions and mergers.

%%%%
The galaxy-galaxy merger timescales ($\sim$10$^9$\,yrs) are substantially longer 
than the ages ($<$10$^5$\,yrs) of radio sources considered here.  Therefore, while 
mergers/interactions may be responsible for funneling large volumes of gas to the 
centers of galaxies they may not be directly involved in triggering the AGN activity.  
This is concurrent with the recent findings that while AGN fraction may be higher amongst 
merging systems, mergers are not the dominant cause of AGN activity \citep[see][]{Ellison11}.  
The AGN activity may actually be triggered by local 
factors such as infall of gas due to disk instabilities. 
%%%
The 21-cm absorption components W$_1$ and W$_2$ detected towards J0942+0623 may be representing 
such gas clouds responsible for feeding the AGN.  
As noted in the previous section, the detection of infalling clumps in absorption is 
not sufficient to conclude that these are also responsible for fuelling the AGN.  
%%%%
Additional constraints on gas geometry and kinematics, possibly from emission lines from ionized and 
molecular gas \citep[e.g.][]{Combes14}, are required to distinguish between various scenarios, 
and understand the role of mergers in triggering the AGN activity.

%%%%%%%%%%%%%%%%%%%%%%%%%%%%%%%%%%%%%%%%%%%
\section{Summary}
%%%%%%%%%%%%%%%%%%%%%%%%%%%%%%%%%%%%%%%%%%%

Using long-slit optical spectroscopy, with the 2-m telescope at IGO, we show 
that SDSS J094221.98+062335.2 is a merging galaxy pair at $z\sim 0.123$ 
with a projected separation of 4.8\,kpc. 
%%%
One of these i.e. object A has radio emission typical of CSOs. 
The measured emission line widths and line flux ratios are consistent with this
object being a narrow line AGN.
The other galaxy, i.e. object B, is found to be a normal star-forming galaxy and has
an infall velocity of  $\sim$185 kms$^{-1}$ with respect to the radio source.

%%%%
We detect strong H~{\sc i} 21-cm absorption associated with the radio source using WSRT and GMRT.  
The \hi column density, $N$(H~{\sc i})=9.1$\times$10$^{21}$(T$_s$/100)(1.0/$f_c$)\,cm$^{-2}$ 
is one of the highest amongst the associated 21-cm absorbers.  
We find that some of the strong 21-cm absorbers with $N$(H~{\sc i})$\sim$10$^{22}$\,cm$^{-2}$ reported in the 
literature are also associated with such major mergers.  
In the WSRT and GMRT spectra, we also detect two well-detached weaker 21-cm absorption components that are 
redshifted with respect to object A by +85 and +195\,\kms. These infalling absorptions could represent gas 
clouds fuelling the AGN.  
All this suggests a possibility that the large volumes of \hi gas detected in these objects are caused 
by merger-driven inflows of gas that {\it could probably be} responsible for triggering and fuelling 
the central AGN.

%%%%
At milliarcsecond scale resolution, the radio source is resolved 
into a CSO like morphology with 88\% of the arcsecond scale flux in two 
dominant continuum components separated by 89\,pc. 
The 21-cm absorption at velocities corresponding to the strong 21-cm absorption component   
detected in the WSRT and GMRT spectra is detected towards both the continuum components 
albeit with an opacity gradient of as much as 3 over 89\,pc.  
The VLBA spectra are not sensitive enough to detect the weaker redshifted absorption components 
mentioned above.
%%%
Comparing the VLBA with the WSRT/GMRT 
spectra we find that the properties of strong 21-cm absorption component are 
consistent with the absorption arising from a circumnuclear disk/torus associated with object A. 
In this model, the opacity gradient across the radio source can be explained by a 
combination of torus half-opening angle and, most importantly, the orientation of 
the line of sight such that the sight line towards the southern VLBA component 
traverses much longer path length through the disk.   
%%%%

%%%%
The available data does not rule out the possibility that instead of originating from gas inflow 
feeding the AGN, the absorptions at +85 and 195\,\kms,  could simply be from (1) intervening 
tidally disrupted gas or (2) gas associated with the ISMs of objects A and B, respectively.   
Additional constraints on gas geometry and kinematics are required to distinguish between various 
scenarios and understand the role of mergers in triggering the AGN activity.

%%%%%%%%%%%%%%%%%%%%%%%%%%%%%% 
\section{acknowledgements}
%%%%%%%%%%%%%%%%%%%%%%%%%%%%%%
We acknowledge the use of SDSS images and spectra from the archive (http://www.sdss.org/).
We thank IGO, GMRT, VLBA and WSRT staff for their support during the observations. 
Special thanks to Dr. Vijay mohan for service observations at IGO. We also
thank the referee Prof. Morganti for useful suggestions.
The VLBA is run by National Radio Astronomy Observatory. The VLBA data
were correlated using NRAO implementation of the DiFX software
correlator \citep{Deller11} that was developed as part of the Australian Major National
Research facilities Programme and operated under license. The National Radio
Astronomy Observatory is a facility of the National Science Foundation operated
under cooperative agreement by Associated Universities, Inc. GMRT is
run by the National Centre for Radio Astrophysics of the Tata Institute of
Fundamental Research.  
%%%%%%%%%%%%%%%%%%%%%%%%%%%%%%%%%%%%%%%
%----------------- Bibliography and bibfile 
%------------------------------------------
\def\aj{AJ}%
\def\actaa{Acta Astron.}% 
\def\araa{ARA\&A}%
\def\apj{ApJ}%
\def\apjl{ApJ}%
\def\apjs{ApJS}%
\def\ao{Appl.~Opt.}%n
\def\apss{Ap\&SS}%
\def\aap{A\&A}%
\def\aapr{A\&A~Rev.}%
\def\aaps{A\&AS}%
\def\azh{AZh}%
\def\baas{BAAS}%
\def\bac{Bull. astr. Inst. Czechosl.}%
\def\caa{Chinese Astron. Astrophys.}%
\def\cjaa{Chinese J. Astron. Astrophys.}%
\def\icarus{Icarus}%
\def\jcap{J. Cosmology Astropart. Phys.}%
\def\jrasc{JRASC}%
\def\mnras{MNRAS}%
\def\memras{MmRAS}%
\def\na{New A}%
\def\nar{New A Rev.}%
\def\pasa{PASA}%
\def\pra{Phys.~Rev.~A}%
\def\prb{Phys.~Rev.~B}%
\def\prc{Phys.~Rev.~C}%
\def\prd{Phys.~Rev.~D}%
\def\pre{Phys.~Rev.~E}%
\def\prl{Phys.~Rev.~Lett.}%
\def\pasp{PASP}%
\def\pasj{PASJ}%
\def\qjras{QJRAS}%
\def\rmxaa{Rev. Mexicana Astron. Astrofis.}%
\def\skytel{S\&T}%
\def\solphys{Sol.~Phys.}%
\def\sovast{Soviet~Ast.}%
\def\ssr{Space~Sci.~Rev.}%
\def\zap{ZAp}%
\def\nat{Nature}%
\def\iaucirc{IAU~Circ.}%
\def\aplett{Astrophys.~Lett.}%
\def\apspr{Astrophys.~Space~Phys.~Res.}%
\def\bain{Bull.~Astron.~Inst.~Netherlands}%
\def\fcp{Fund.~Cosmic~Phys.}%
\def\gca{Geochim.~Cosmochim.~Acta}%
\def\grl{Geophys.~Res.~Lett.}%
\def\jcp{J.~Chem.~Phys.}%
\def\jgr{J.~Geophys.~Res.}%
\def\jqsrt{J.~Quant.~Spec.~Radiat.~Transf.}%
\def\memsai{Mem.~Soc.~Astron.~Italiana}%
\def\nphysa{Nucl.~Phys.~A}%
\def\physrep{Phys.~Rep.}%
\def\physscr{Phys.~Scr}%
\def\planss{Planet.~Space~Sci.}%
\def\procspie{Proc.~SPIE}%
\let\astap=\aap
\let\apjlett=\apjl
\let\apjsupp=\apjs
\let\applopt=\ao
\bibliographystyle{mn}
%\bibliography{/Users/ngupta/Desktop/mybib}
\bibliography{../../mybib1}

\end{document}